\documentclass[prl,twocolumn,showpacs,amsmath,superscriptaddress,amssymb]{revtex4}

\usepackage{graphicx}
\usepackage{dcolumn}
\usepackage{bm}

\begin{document}


\title{Gapless interface states between topological insulators with opposite Dirac 
velocities}

\author{Ryuji Takahashi}
 \affiliation{Department of Physics, Tokyo Institute of Technology,
2-12-1 Ookayama, Meguro-ku, Tokyo 152-8551, Japan}

\author{Shuichi Murakami}
\affiliation{Department of Physics, Tokyo Institute of Technology, 
2-12-1 Ookayama, Meguro-ku, Tokyo 152-8551, Japan} 
\affiliation{PRESTO, Japan Science and Technology Agency (JST),
Kawaguchi, Saitama 332-0012, Japan} 

\date{\today}

\begin{abstract}
The Dirac cone on a surface of a topological insulator shows linear dispersion analogous to optics
and its velocity  
depends on materials.
We consider a junction of two topological insulators with different velocities, and 
calculate
the reflectance and transmittance. 
We find that 
they reflect the backscattering-free nature of the 
helical surface states.  
When the two velocities 
have opposite signs, both 
transmission and reflection are prohibited for normal incidence, when a mirror symmetry normal to the 
junction is
preserved.
In this case we show that there necessarily exist gapless states at the interface between the two topological insulators. 
Their existence is protected by mirror symmetry, and they have characteristic 
dispersions depending on the symmetry of the system.
\end{abstract}

\maketitle
Recently 
physical phenomena originating from the Dirac cones of electrons have been studied,
in the context of graphene sheet \cite{Novoselov04} or the topological insulator (TI) 
\cite{Kane05a,Bernevig06,Moore07,Fu07}.
In a graphene sheet, novel transport phenomena are predicted theoretically 
in $p$-$n$ junction systems: for example the Klein paradox \cite{Katsnelson06},
and the negative refraction \cite{Cheianov07}. The TI in three dimensions (3D) \cite{Fu07,Moore07},
such as Bi$_2$Se$_3$ \cite{Xia09,Hsieh09} and Bi$_2$Te$_3$ \cite{Chen09}, 
has a single Dirac cone in its surface states, as observed by 
angle-resolved
photoemission spectroscopy.
Unlike graphene, the states on the Dirac cone on the surface of the TI are spin filtered; they have fixed spin directions for each wave number ${\bf k}$.  
Because the state at ${\bf k}$ and that at $-{\bf k}$ have the opposite spins, the perfect backscattering
from ${\bf k}$ to $-{\bf k}$ is forbidden.

\begin{figure}[htbp]
 \begin{center}
  \includegraphics[width=85mm]{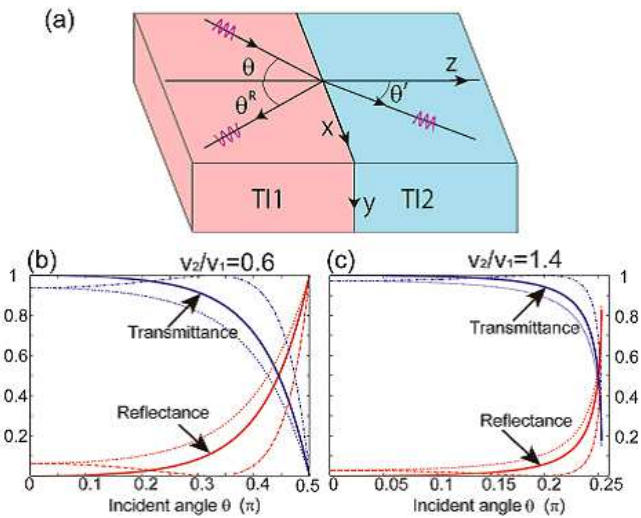}
 \caption{(Color online) (a) Schematic of the refraction of the surface states
at the junction between the two TIs, TI1 and TI2. 
(b)(c): Reflectance (red) and transmittance (blue) for 
 the ratios of the velocities of the two TIs: (b)$v_{2}/v_{1}=0.6$ and (c)$v_{2}/v_{1}=1.4$.
The solid curves are the results for the junction between two TIs, while the dotted curves show
the results for
 optics with $p$ and $s$ polarizations.
}
\label{fig:kussetsu}
\end{center}\end{figure}

Such linear dispersion is similar to photons.
The velocity of the Dirac cone on the surface of 3D TI depends on materials.
 For example, the velocity for Bi$_2$Te$_3$ is about $4\times10^5$m/s \cite{Chen09} depending on the direction of the wave vector, and that for Bi$_2$Se$_3$ 
is approximately
$5\times10^5$ m/s \cite{Xia09}.  
Therefore, when two different TIs are attached together, the refraction phenomenon similar to  
optics is expected at
the junction. In this Letter, we theoretically study the refraction of electrons at the junction
between 
the surfaces of two TIs 
[Fig.~\ref{fig:kussetsu}(a)]. The resulting transmittance and reflectance are different from optics,
 reflecting prohibited perfect backscattering.
In addition, we show that  
when the velocities of the two TIs have opposite signs, neither refraction nor reflection is allowed for
the incident electron normal to the junction.  
In this case, we can show that there necessarily exist gapless interface states between the two TIs and the incident surface electrons totally go into the interface states.
As long as the mirror symmetry with respect to the $yz$ plane ${\cal M}_{yz}$ is preserved, the interface gapless states exist.
These gapless 
states are formed at the interface between the same Z$_2$ nontrivial materials.
As a result, these
interface states do not come from the Z$_2$ topological number, but come from the mirror Chern number\cite{Teo08}, 
and are protected by the mirror symmetry ${\cal M}_{yz}$.

The effective Dirac Hamiltonian of the surface states on the $xz$ plane is represented as
\begin{eqnarray}
H&=&-iv[\sigma_{x}\partial_{z}-\sigma_{y}\partial_{x}],\label{RHamiltonian}
\end{eqnarray}
where $\sigma_x$, $\sigma_y$ are the Pauli matrices, and $v$ is the Fermi velocity.
From the Hamiltonian one can obtain the linear energy $E=svk$ where $k=|\mathbf{k}|$, and $s=+1(-1)$ corresponds to the upper (lower) cone, provided $v>0$. 
We consider a refraction problem between the two TIs, which we call TI1 and TI2, with
 the incidence angle $\theta$, the transmission
angle $\theta'$, and the reflection angle $\theta^R$ [Fig.~\ref{fig:kussetsu}(a)].
As in optics, the momentum conservation requires $\theta^{R}=\theta$, and 
the wave functions are written as
$\psi^{I}(x,z)=\frac{1}{\sqrt{2}}\mathrm{e}^{ik(x\sin\theta+z\cos\theta)}
(1,\mathrm{e}^{-i\theta})^{t},\ 	
\psi^{T}(x,z)=\frac{1}{\sqrt{2}}\mathrm{e}^{ik'(x\sin\theta'+z\cos\theta')}
(1,\mathrm{e}^{-i\theta'})^{t},\ 
\psi^{R}(x,z)=\frac{1}{\sqrt{2}}\mathrm{e}^{ik(x\sin\theta-z\cos\theta)}
(1,-\mathrm{e}^{i\theta})^{t},
$
where $k$ and $k'$ are the wave numbers on TI1 and TI2, respectively, and we consider the Fermi energy $E_{F} >0$ (i.e above the Dirac point), giving $s=+1$ for both of the TIs.
Let $v_1$ and $v_2$ denote the velocities of the two TIs. 

We first assume $v_1$ and $v_2$ to be positive. The conservation of the momentum and the energy yields  Snell's law: 
$k'\sin\theta'=k\sin\theta, \ 
v_1^{-1}\sin\theta=v_2^{-1}\sin\theta'$.
Let $r$ and $t$ denote the amplitude of the reflected and transmitted wave, compared with the incident wave. 
The current conservation in the $z$ direction 
is written as $R+T=1$, where $R\equiv
|r|^2$ and $T\equiv\frac{v_{2}\cos\theta'}{v_{1}\cos\theta}|t|^2$ are
the reflectance and the transmittance, respectively.
We note that the wavefunction should eventually be discontinuous at the junction when the
 velocities are different, as has been studied in the context of graphene \cite{Raoux10,Concha10}. The reason is the following. 
Therefore, the current conservation
 at the interface requires $v_{1z}|\psi_{1}|^2=v_{2z}|\psi_{2}|^2$, 
where ${\bf v}_{1,2}$ is a velocity, and the subscripts 1 and 2 represent TI1 and TI2, respectively. 
Because in our case $v_{1z}\neq v_{2z}$, we have $|\psi_{1}|^2\neq|\psi_{2}|^2$ at the junction, and the continuity of the wavefunction is violated. The proper 
way 
is to set the Hamiltonian to be Hermitian also at the boundary, i.e. 
$H=-i\left[\frac{1}{2}[v(z)\sigma_{x}\partial_{z}+\sigma_{x}\partial_{z}v(z)]-v(z)\sigma_{y}\partial_{x}\right],$ where $v(z)$ is the velocity dependent on $z$.
The resulting coefficients are
\begin{eqnarray}
r=i\frac{\sin\frac{\theta'-\theta}{2}}{\cos\frac{\theta+\theta'}{2}}\mathrm{e}^{-i\theta}, \ \ 
t=\sqrt{\frac{v_1}{v_2}}\frac{\cos\theta}{\cos\frac{\theta+\theta'}{2}}\mathrm{e}^{i\frac{\theta'-\theta}{2}}.
\label{eq:coefficisnts}
\end{eqnarray}
They satisfy the current conservation.
The results are plotted as the solid curves in Figs.~\ref{fig:kussetsu}(b)(c). The dotted curves represent corresponding results for
optics. 
Unlike optics, for normal incidence ($\theta=0$), 
the perfect transmission ($T=1$, $R=0$) occurs, 
which reflects the prohibited backscattering on the surface of the TI.
This is similar to graphene \cite{Katsnelson06,Raoux10,Concha10} but the transmittance in our
case monotonically decreases
with the incidence angle.

\begin{figure}[htbp]
 \begin{center}
  \includegraphics[width=80mm]{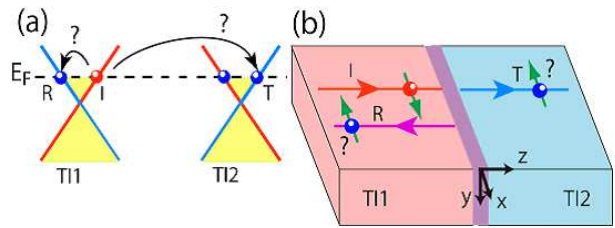}
 \end{center}
 \caption{(Color online) Transport at the junction between the surfaces of two TIs,
 whose velocities have different signs. (a) Linear dispersion at $k_{x}=0$. 
The incident wave (I) is perpendicular to the junction.
Both the transmission (T) and reflection (R)
are prohibited due to spin conservation. (b) 
Normal incidence. TI1 (red) and TI2 (blue) have the velocities of opposite signs. The purple region represents the interface.}
 \label{fig:TIjunction}
\end{figure}

Next, we consider the case where the velocities of the two TIs have opposite signs,
where we can no longer use the above approach.
One might think that it is similar to the negative refraction in optics \cite{Veselago,Pendry},
but it is not true because the Fermi energy is above the Dirac point for the 
two TIs. 
Furthermore, 
both reflection and transmission are prohibited for normal incidence, 
because
the incident wave has no way to conserve its momentum $k_x$ along the interface and spin simultaneously (see Fig.~\ref{fig:TIjunction}).
Thus it is a paradox what happens for normal incidence.

Our answer to this question is that 
gapless states exist at the interface between the two TIs
(the purple  region in Fig.~\ref{fig:TIjunction}(b)).
The normally incident wave goes along the surface of one TI, then into the interface between the two TIs.
These interface states arise from hybridization between
the two surface states from the two TIs.
To show the existence of gapless interface states, we first 
write down the effective Hamiltonian at 
the interface from the two Dirac cones with 
hybridization:
\begin{eqnarray}
H=
\begin{pmatrix}
H_{1}&V\\
V^{\dagger}&H_{2}
\end{pmatrix}. \label{eq:mhamiltonian}
\end{eqnarray}
Here 
$H_{1(2)}$ is the effective surface Hamiltonian for the
surface of TI1 (TI2) at the interface:
\begin{eqnarray}
H_{1}=v_{1}( \bm{\sigma}\times{\bf k})_z
, \ 
H_{2}=-v_{2}(\bm{\sigma}\times{\bf k})_z
\label{H1H2}
\end{eqnarray}
and $V$
is the hybridization at the interface.
For simplicity, we retain only the lowest order in ${\bf k}$.
In the expression of $H_2$, there is an extra minus sign; on the surface of TI2 in Fig.~\ref{fig:TIjunction}, the mode going in the $+z$ direction evolves from
that going in the $-y$ 
direction, whereby the extra sign necessarily appears. 

We explain the reason for justifying our model in Eqs.~(\ref{eq:mhamiltonian}), (\ref{H1H2}). 
For simplicity we assumed that the surface states on the $xz$ surface for TI1 and TI2 are described by the Dirac cone. Generic surface states with non-Dirac types are covered in the later
discussion using the mirror Chern number\cite{Teo08}. 
We used here the fact that the Dirac velocities for each TI have the same signs for the $xy$ and $xz$ surfaces. It is because the signs of the Dirac velocities are determined by 
the mirror Chern number which is the bulk quantity \cite{Teo08}. 
We also set the Dirac cones to be isotropic for simplicity; the following 
results turn out to be unaltered by anisotropy in 
the Dirac cones.
We henceforth impose the mirror symmetry with respect to the $yz$ plane ${\cal M}_{yz}$, because this symmetry  
preserved by $H_1$ and $H_2$ sets the spins parallel to the $x$ axis for 
the normally ($\|\hat{z}$) incident wave. 
By imposing this mirror symmetry ${\cal M}_{yz}$ and time-reversal symmetry, $V$ is expressed as 
$
V=
\begin{pmatrix}
g&ih\\
ih&g
\end{pmatrix},
$
where
$g$ and $h$ are real constants representing 
the hybridization between the two surface states.
From the Hamiltonian [Eq.~(\ref{eq:mhamiltonian})], the eigenvalues are calculated as
\begin{eqnarray}
E&=&\pm\sqrt{\Delta_k\pm
\sqrt{{\Delta}^2_k-\eta}
},\ 
\Delta_k=g^2+h^2+\frac{v^2_{1}+v^2_{2}}{2}k^2,\ 
\label{eq:interfaceenergy}\\
\eta&=&v^2_{1}v^2_{2}k^4+\Delta_0^2-2v_{1}v_{2}k^2 (h^2\cos 2\alpha
-g^2), 
\end{eqnarray}
where we set $k_{x}+ik_{y}= k\mathrm{e}^{i\alpha}$, and $\alpha$ is real.
The condition for existence of gapless interface states is
\begin{eqnarray}
v^2_{1}v^2_{2}k^4-2v_{1}v_{2}k^2 (h^2\cos 2\alpha
-g^2)+(g^2+h^2)^2=0.
\label{eq:gaplesscond}
\end{eqnarray}
To solve this equation, we note that $g$ is nonzero, whereas $h$ can become
zero when additional symmetries such as rotational symmetry with respect to the $z$ axis are
imposed. 
Then we can see that
for $v_1v_2>0$ (the two velocities with the same signs),  the interface states 
are gapped by the hybridization. 

\begin{figure}[t]
\begin{center}
  \includegraphics[width=85mm]{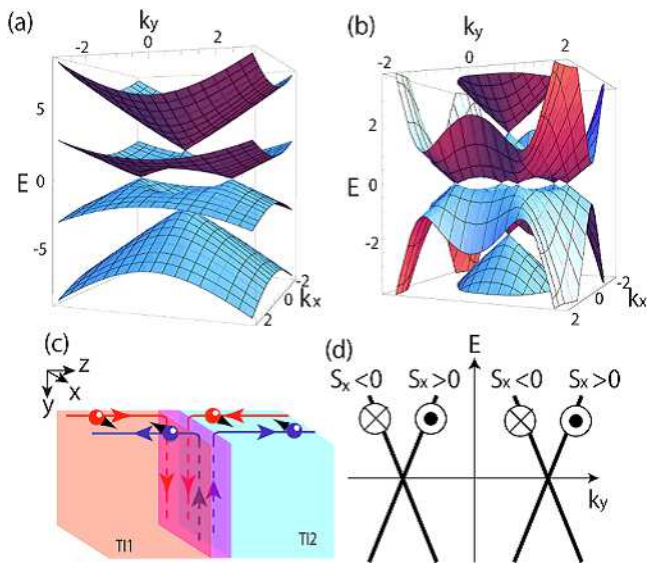}
 \end{center}
 \caption{(Color online) (a)(b) Dispersion on the interface between the two TIs in Eq.~(\ref{eq:interfaceenergy}) with velocities $v_1=1$, $v_2=-2$.
 In (a), the hybridization is $g=2$, $h=1$. There are two Dirac points
  $(k_{x}, k_{y})=(0,\pm\sqrt{5/2})$ where the gap closes. 
In (b), the hybridization is $g=2$, $h=0$ and the warping term 
$\lambda(k_+^3+k_-^3)\sigma_z$ with $\lambda=0.4$ added to $H_1$ and $H_2$. 
There appear six Dirac cones.
(c) Illustration of the interface mode. The surface current goes into the interface.
(d) Schematic of the dispersion of interface states on $k_x=0$.}
 \label{fig:gaplessmode}
\end{figure}

Only when two velocities have opposite signs
($v_1v_2<0$), are there gapless states on the interface. 
Dispersion of the gapless states depends on whether $h\neq 0$ or $h=0$. 
When $h\neq 0$,
the solutions are
$(k_{x},k_{y})=\left(0,\pm\sqrt{(g^2+h^2)/|v_{1}v_{2}|}\right)$
and there are gapless states on the interface. 
The interface states have two Dirac cones (Fig.~\ref{fig:gaplessmode}
(a)).  
On the other hand, when $h=0$ due to rotational symmetry with respect to 
the $z$ axis, the gap closing points form a circle
$k_{x}^2+k_{y}^2=\frac{g^2}{|v_{1}v_{2}|}$.
This degeneracy on the circle in ${\bf k}$ space is due to the continuous rotational 
symmetry around the $z$ axis, and is lifted when it is broken by adding higher order terms
in ${\bf k}$, e.g. the warping term in Bi$_2$Se$_3$ \cite{Fu09}. 
As seen in Fig.~\ref{fig:gaplessmode}(b),
the dispersion becomes a collection of Dirac cones.
Therefore, for this example Hamiltonian, we could show that there are gapless states at the interface when the system has the mirror symmetry ${\cal M}_{yz}$.

We note that this method is generic, because the analysis is based only on the symmetry. 
The only assumption is that the gapless point is near ${\bf k}=0$, and we can expand the 
Hamiltonian in terms of $k$.
To complement this argument, we show the existence of gapless interface states on generic grounds. 
Because these gapless states are generated between two TIs with the same 
Z$_2$ topological numbers, they are not protected in the same sense as the surface states of three-dimensional
TIs. 
In the following we show 
that these gapless interface states are protected by the mirror symmetry and the time-reversal symmetry.
Each TI with mirror symmetry is characterized by the mirror Chern number \cite{Teo08}.
When the system has the mirror symmetry ${\cal M}_{yz}$,
 the surface modes are labeled with the mirror eigenvalues $\mathcal{M}=\pm i$ at $k_x = 0$, 
corresponding to the spin along $-x$ and $+x$, respectively. The mirror Chern number
 is obtained as
$n_{\mathcal{M}}=(n_{+i}-n_{-i})/2$
where $n_{\pm i}$ are the Chern numbers \cite{Thouless1982,
Kohmoto1985
} for the subspace
of states 
with mirror eigenvalues $\mathcal{M}=\pm i$. 
We have $n_{+i}=-n_{-i}$ by the time-reversal symmetry.
In our case where the two surface Dirac cones have opposite velocities, the mirror Chern numbers for the two TIs are different. TI1 has $n_{\mathcal{M}}^{(1)}=-1$, i.e. $n_{+i}^{(1)}=-1, n_{-i}^{(1)}=+1$, and TI2 has $n_{\mathcal{M}}^{(2)}=1$, i.e. $n_{+i}^{(2)}=+1, n_{-i}^{(2)}=-1$ at $k_{x}=0$ plane. 
For the $\mathcal{M}=+i$ ($S_x<0$) subspace, this corresponds to the junction of two systems with Chern numbers $n_{+i}^{(1)}=-1$ and $n_{+i}^{(2)}=+1$;
because $n_{+i}^{(1)}-n_{+i}^{(2)}=-2$, it gives rise to two left-going chiral modes in the $y$ direction. 
On the other hand, for $\mathcal{M}=-i$ it also gives two right-going chiral modes in the $y$ 
direction. 
These modes are schematically shown in Fig.~\ref{fig:gaplessmode}(d).
Therefore it is natural to generate the two Dirac cones in the junction.
Thus these gapless states are protected by the mirror symmetry.
If the mirror symmetry is not preserved, the gapless states do not exist in general.
This discussion is 
generic, and is complementary to our discussion by the surface Dirac Hamiltonian. 
Therefore, we conclude that the gapless interface states exist
for the generic cases with mirror symmetry, even with e.g. lattice mismatch at the interface.
In real materials, mirror symmetry may be lost by disorder in principle; nevertheless, if the 
the sample is relatively clean, the gapless interface states are expected to survive and can be
measured experimentally. 

The distance between the Dirac cones of the gapless interface states are proportional 
to the magnitude of the hybridization between the two TIs at the interface. 
When the hybridization becomes as strong as the 
bandwidth, the spacing between the interface Dirac cones is of the order of inverse of the lattice 
spacing. In that case the transport properties will be like the graphene, having two Dirac cones at K and K' points. 
We note that in graphene there are spin-degenerate Dirac cones, whereas 
in the present case the interface Dirac cones are not spin degenerate.
From Fig.~\ref{fig:gaplessmode}(d), 
when the wave number ${\bf k}$ goes around one of the 
Dirac point, the spin direction also rotates around the $z$ axis (normal to the interface). 
In the similar way as in graphene, one can consider the valley degree of freedom as a pseudospin, and 
develop valleytronics \cite{Gunawan2006,Xiao2007} similar to graphene. 
These interface states can be measured via transport; for this purpose one should suppress the surface 
transport by attaching ferromagnets on the surface.

From the spin-resolved angle-resolved photoemission spectra, 
all the TIs observed so far, such as Bi$_{1-x}$Sb$_x$ \cite{Hsieh09a,Nishide}, Bi$_2$Se$_3$\cite{Hsieh09}, 
and Bi$_2$Te$_3$\cite{Hsieh09}, 
have $n_{\cal M}=-1$. 
To realize the protected interface states in experiments discussed in this Letter, 
one needs to find a TI with $n_{\cal M}=+1$,
i.e., the surface Dirac cone with negative velocity, and the 
spins on the upper cone is in the counterclockwise direction in the ${\bf k}$ space.
It is an interesting issue to search for such TIs. 
The Dirac velocity $v$ is nothing but the coefficient $\lambda$ in the 
Rashba spin-splitting term $\lambda(\bm{\sigma}\times{\bf k})_z$ in the Hamiltonian.
The  Rashba coefficient $\lambda$ originates from  
an integral of a sharply peaked function near the nuclei, which rapidly varies between positive 
and negative values
\cite{Bihlmayer,Nagano}.
Therefore, we expect that it can change sign in principle.
The sign of the mirror Chern number $n_{\cal M}$ is also related with the mirror chirality 
of the bulk Dirac Hamiltonian describing the bands near the bulk gap \cite{Teo08}.
Because the mirror chirality 
governs the sign of the $g$-factor which can be negative or positive as
a result of the spin-orbit coupling, one may well expect that in some materials 
$n_{\cal M}$ can become $+1$.

In conclusion, we study refraction phenomena on the junction between the two TI surfaces with  different velocities. 
The resulting
 reflectance and transmittance 
reflect the backscattering-free nature of the surface states of TIs. 
When the velocities of the TI surface states for the two TIs have different signs, 
we show that the gapless states appear on the interface. The existence of the  gapless states is shown by using the mirror Chern number, and thus is topologically protected by the mirror symmetry.

The authors are grateful to T. Oguchi for discussions.
This research is supported in part 
by Grant-in-Aid for Scientific Research (No.~21000004 and 22540327)
from the MEXT, Japan
and by Kurata Grant from Kurata Memorial Hitachi Science and Technology Foundation.
R.T also acknowledges the financial support from the Global
Center of Excellence Program by MEXT, Japan through the
``Nanoscience and Quantum Physics" Project of Tokyo Institute of Technology.

%


\end{document}